\documentclass[onecolumn]{pasj01}

\begin{document} 
\Received{2017/06/27}
\Accepted{2018/01/25}

\title{Molecular clouds in the {NGC 6334} and {NGC 6357} region;
Evidence for a {100 pc}--scale cloud-cloud collision triggering the Galactic mini-starbursts }

\author{Yasuo \textsc{Fukui}\altaffilmark{1,2}$^{*}$}%
\altaffiltext{1}{Department of Physics, Nagoya University, Furo-cho, Chikusa-ku, Nagoya, Aichi 464-8601, Japan}
\altaffiltext{2}{Institute for Advanced Research, Nagoya University, Furo-cho, Chikusa-ku, Nagoya 464-8601, Japan}
\altaffiltext{3}{Nobeyama Radio Observatory, National Astronomical Observatory of Japan (NAOJ), National Institutes of Natural Sciences (NINS), 462-2, Nobeyama, Minamimaki, Minamisaku, Nagano 384-1305, Japan}
\email{fukui@a.phys.nagoya-u.ac.jp}
\author{Mikito \textsc{Kohno}\altaffilmark{1}$^{*}$}
\email{mikito@a.phys.nagoya-u.ac.jp}
\author{Keiko \textsc{Yokoyama}\altaffilmark{1}}
\author{Kazufumi \textsc{Torii}\altaffilmark{3}}
\author{Yusuke \textsc{Hattori}\altaffilmark{1}}
\author{Hidetoshi \textsc{Sano}\altaffilmark{1,2}}
\author{Atsushi \textsc{Nishimura}\altaffilmark{1}}
\author{Akio \textsc{Ohama}\altaffilmark{1}}
\author{Hiroaki \textsc{Yamamoto}\altaffilmark{1}}
\author{Kengo \textsc{Tachihara}\altaffilmark{1}}



\maketitle

\begin{abstract}
We carried out new CO ($J=$1--0, 2--1 and 3--2) observations with NANTEN2 and ASTE in the region of the twin Galactic mini-starbursts {NGC 6334} and {NGC 6357}. We detected two velocity molecular components of 12 km s$^{-1}$ velocity separation, which is continuous over 3 degrees along the plane. In {NGC 6334} the two components show similar two-peaked intensity distributions toward the young H\,\emissiontype{II} regions and are linked by a bridge feature. In {NGC 6357} we found spatially complementary distribution between the two velocity components as well as a bridge feature in velocity. Based on these results we hypothesize that the two clouds in the two regions collided with each other in the past few Myr and triggered formation of the starbursts over $\sim$100 pc. We suggest that the formation of the starbursts happened toward the collisional region of $\sim$ 10-pc extents with initial high molecular column densities. For {NGC 6334} we present a scenario which includes spatial variation of the colliding epoch due to non-uniform cloud separation. The scenario {possibly} explains the apparent age difference among the young O stars in {NGC 6334} raging from 10$^4$ yrs to 10$^6$ yrs; the latest collision happened within 10$^5$ yrs toward the youngest stars in {NGC 6334} {I(N) and I} which exhibit molecular outflows without H\,\emissiontype{II} regions. For {NGC 6357} the O stars were formed a few Myrs ago, and the cloud dispersal {by} the O stars is significant. We conclude that cloud-cloud collision offers a possible explanation of the min-starburst over a 100-pc scale.
\end{abstract}

\section{Introduction}
\subsection{Background}

High-mass stars are extremely influential in galaxy evolution by dynamically agitating the interstellar medium (ISM) and by injecting heavy elements, and, in particular, O stars whose mass exceeds 20 $M_{\odot}$ are most influential. It is, therefore, one of the most important issues to understand formation of high-mass stars in galaxies. Infrared dark clouds (IRDCs) having extremely high molecular column density are discussed as one of the promising sites of high-mass star formation. This is a natural direction {in pursuing} the origin of high-mass stars that requires massive dense cores. This seems to be supported by the monolithic collapse picture theoretically {(McKee \& Tan 2003, Krumholz et al. 2009)}. It is, however, puzzling why we have no direct evidence for real high-mass star formation of developed H\,\emissiontype{II} regions in {many of} the IRDCs {(e.g., Peretto et al. 2014)}. Another issue to be addressed is that the mass accretion rate has to be very high, $\sim 10^{-4}- 10^{-3}$ $M_{\odot}$ yr$^{-1}$, to form O stars with mass greater than 20 $M_{\odot}$ (Wolfire \& Cassinelli 1987). This condition requires a high effective sound speed because the mass accretion rate is proportional to the third power of the sound speed; {which is not {well} supported by observed molecular linewidths {of a few km s$^{-1}$ in IRDCs ({e.g., Henshaw et al. 2013}).}

In the past several years there is increasing evidence for cloud-cloud collision as a trigger of O star formation in the Milky Way and the LMC, and the number of young O stars with colliding clouds are more than 20 up to now; Sgr B2 (Hasegawa et al. 1994), Westerlund2 (Furukawa et al, 2009; Ohama et al. 2010), NGC3603 (Fukui et al. 2014), RCW38 (Fukui et al. 2016), DBS[2003]179 (Kuwahara et al. in preparation), M20 (Torii et al. 2011), RCW120 (Torii et al. 2015), N37 (Baug et al. 2016), G35.20-0.74 (Dewangan 2017), L1188 (Gong et al. 2017), {Galactic center 50 km s$^{-1}$ molecular cloud} (Tsuboi et al. 2015), M42 and M43 (Fukui et al. {2018a}), W51 (Fujita et al. {2018}), {M16 (Nishimura et al. {2018b}), {W33 (Kohno et al. {2018})}, M17 (Nishimura et al. {2018a}), GM 24 (Fukui et al. 2017c), RCW34 (Hayashi et al. {2018}), RCW36 (Sano et al. {2018}), NGC2024 (Ohama et al. {2018a}), RCW166 (Ohama et al. {2018b}), N159W and N159E (Fukui et al. 2015; Saigo et al. 2016), R136 (Fukui et al. 2017a), and N44 (Tsuge et al. in preparation). These observations show that O stars or early B stars are associated with two velocity gas components which show signatures of mutual collisional interaction such as bridge features and complementary distributions. As such, it is becoming important to explore if {cloud-cloud collision} plays an important role in forming high-mass stars, {and to examine} if {cloud-cloud collision} is an important mechanism of O star formation.

Theoretical studies by Inoue and Fukui (2013) have shown collisions between two molecular clouds at a {supersonic} velocity can excite turbulence and {amplify} magnetic field in the shock-compressed interface layer leading to form dense and massive cores which {form O stars}. The enhanced sound speed realizes a high-mass accretion rate of $10^{-4} M_{\odot}$ yr$^{-1}$ to $10^{-3} M_{\odot}$ yr$^{-1}$ which satisfies the requirement to form high-mass stars by overcoming the stellar radiation feedback. It is therefore without doubt that cloud-cloud collision provides physical conditions favorable for high-mass star formation, whereas it is not established if cloud-cloud collision is a major mechanism for high-mass star formation. 
A concern is if cloud-cloud collision, a seemingly ad-hoc event, is frequent enough to explain the majority of high-mass star formation. A weakness of IRDCs as the O star formation site lies in that the cloud motion is too quiet to achieve the high-mass accretion rate.
Theoretical studies show that high-mass stars can be formed without trigger (Krumholz et al. 2009), whereas the initial conditions assumed remain to be verified {to be} attainable without trigger through organized observational studies of dense cores. In order to better understand and prove the origin of high-mass stars, it is crucial to obtain molecular gas data with high quality, high sensitivity, high resolution, and large spatial coverage, in various kinds of H\,\emissiontype{II} regions, unmistakable high-mass star formation regions.

\subsection{Twin starbursts in the Carina--Sagittarius arm}

{NGC 6334} and {NGC 6357} are remarkable Galactic high-mass star formation regions {known as} “twin mini starbursts”, which include more than 30 O stars, most of which have H\,\emissiontype{II} bubbles or compact H II regions (see for a review Persi \& Tapia 2008, hereafter PT08, and references therein). It is probable that more OB stars are embedded in the region which remain undiscovered due to heavy extinction. The region is very complicated with a number of gas/dust condensations, as well as mixture of high- to low-mass stars. Figures 1 and 2 show {the Herschel infrared image at 70, 250, and 500 $\mu$m (Molinari et al. 2010, {Pilbratt et al. 2010; Poglitsch et al. 2010; Griffin et al. 2010}) and $^{12}$CO $J=$ 1--0 distribution}. The two starburst regions are separated by {$\sim$2 degrees} along the {Galactic} plane. 

{NGC 6334} is located at a distance of 1.61 kpc (PT08), and consists of a dense molecular gas filament extended along the Galactic plane over 10 pc and extended H\,\emissiontype{II} regions outside the filament. The filamentary molecular ridge is dense and consists of six active sub-regions identified by McBreen et al. (1979) and Loughran et al. (1978) as far-infrared sources which are denominated I-V and I(N) (see Table 3 in TP08, and recent papers Tig\'e et al. 2017; And\'re et al. 2016; Russeil et al. 2010, 2012, 2013, 2016; Willis et al. 2013, Zernickel et al. 2013; Feigelson et al. 2009). {There are molecular outflow{s} associated {with {I(N) and I}} (Megeath \& Tieftrunk,1999, McCutcheon et al. 2000, Leurini et al. 2006, Qiu et al. 2011)}. Each of these sub-regions includes at least a few high-mass stars whose spectral types are {from B0 to O6.5}, and are associated with H$_2$O, OH, NH$_3$ and CH$_3$OH masers as well as $\sim$ 100 low-mass young stars with extents from 1 to several acmin. It is suggested that ages of these far-infrared sources range from $10^4$ yrs to $10^6$ yrs {as listed in Table 1, where the ages are taken from Table 3 of PT08}. {Outside the} regions of heavy extinction, optical diffuse nebulae are extended where the bubble-like H II regions including GUM 61, GUM 62, GUM 63, GUM 64, and H II 351.2+0.5 are located alongside of a nebula GM 24 in the west, which are ionized by $\sim$ 10 OB stars (for other nomenclature{s} see Table 1 of TP08), and are older than the sub-regions in the filament.
{NGC 6357} is another nearby region of extreme high-mass star formation. In particular, Pismis 24-1 (HD 319718) is a multiple O star including a spectroscopic binary with an O3.5 supergiant and an O4 giant star. The nebula accompanies extended H\,\emissiontype{II} regions similar to {NGC 6334}, whereas its central region is {bubble-like} with less obscuration than in {NGC 6334}. After some controversy it seems now widely accepted that {NGC 6334} and {NGC} 6357 are both located at a distance of $\sim$1.7 kpc ({Massi et al. 2015)}.

There remain many issues to be addressed in the {cloud-cloud collision} scenario. One of them is the difference in the O star distribution in {cloud-cloud collision}; the super star clusters show very compact O star distribution less than 1 pc. The observations of RCW38 {showed} that the {shape and} size of the O star {distribution} {are} possibly determined by the area of the collisional interaction (Fukui et al. 2016).  The collision in M42 also suggests that the O star distribution is constrained by the size of collision (Fukui et al. {2018{a}}). It is noted that there is a threshold of molecular column density for formation of more than ten O stars (Fukui et al. {2018a}). Considering the spatially extended nature of O star formation in {NGC 6334} and {NGC 6357}, one may expect to obtain a {pivotal clue} to better understand the O star distribution in these objects. 

In order to reveal large-scale distribution of the parental molecular gas of the two regions and to shed a new light on the formation of the two starbursts, we carried out new observations of CO with NANTEN2 and ASTE. Section 2 describes observations and Section 3 NANTEN2 and ASTE results. Section 4 gives discussion on cloud-cloud collision and Section 5 concludes the paper.

\section{Observations}

We performed $^{12}$CO ($J=$ 1--0, 2--1) observations with the NANTEN2 4 m millimeter/sub-millimeter radio telescope of Nagoya University. The observations of $^{12}$CO ($J =$ 1--0) emission were conducted from May 2012 to December 2012. The front end was a 4 K cooled SIS mixer receiver. The system temperature including the atmosphere was $\sim$ 120 K in the double-side band (DSB) including the atmosphere toward the zenith. 
The backend was a digital-Fourier transform spectrometer (DFS) with 16384 channels of 1 GHz bandwidth. The velocity coverage and resolution were $\sim 2600$ km s$^{-1}$ and 0.16 km s$^{-1}$, respectively. We used the on-the-fly (OTF) mapping mode, and the observed area was $\sim$ \timeform{3.5D} $\times$ \timeform{0.8D}. The pointing accuracy was confirmed to be better than \timeform{20"} with daily observations toward the Sun. 
The absolute intensity calibration was applied by observing IRAS 16293−2422 [$\alpha_{\rm J2000} = \timeform{16h32m23.3s} , \delta_{\rm J2000} = \timeform{-24d28'39. 2’’}$] (Ridge et al. 2006). 
{The data cube was smoothed with a Gaussian kernel of \timeform{90"}, and the final beam size was \timeform{180"} (FWHM).} The typical rms noise level was $\sim$1.2 K ch$^{-1}$ {with a velocity resolution of 0.16 km s$^{-1}$}. 

  Observations of $^{12}$CO ($J =$ 2--1) emission were carried out from October 2014 to November 2015. The typical system temperature was $\sim180$ K  {in the double side band} including the atmosphere {toward} the zenith. We observed molecular clouds toward {NGC 6334} and {NGC 6357} with the OTF mapping mode. The observed area {is} shown in Figure 2. The velocity coverage and resolution were $\sim 1300$ km s$^{-1}$ and 0.08 km s$^{-1}$, respectively. {Pointing} accuracy {of} $\sim$ \timeform{10"} was confirmed by observing the Moon. The absolute intensity was calibrated by observing {M17SW[$\alpha_{\rm J2000} =\timeform{18h20m24.4s}, \delta_{\rm J2000} = \timeform{-16D13'17.6"}$]}. The data cube was smoothed with a Gaussian kernel of \timeform{45"}, and the final beam size was \timeform{90"} (FWHM). The typical rms noise level was $\sim 1.1$ K ch$^{-1}$ {with a velocity resolution of 0.08 km s$^{-1}$}.
  
  Subsequently, we carried out $^{12}$CO ($J=$ 3--2) and $^{13}$CO ($J=$ 3--2) observations toward the central region of {NGC 6334} with the Atacama {Sub-millimeter Telescope Experiment (ASTE, Ezawa et al. 2004, 2008) from July to September 2014. The frontend was 2SB SIS mixer socalled “CATS 345 ” (Inoue et al. 2008).} The typical system temperature was {350 K at 345 GHz} in the single-side band (SSB). The {backend} was the DFS, “MAC” (Sorai et al. 2000), with 1024 channels of 128 MHz bandwidth. We used the OTF mapping mode and {an} observed area was \timeform{0.43D} $\times$ \timeform{0.14D}. {Pointing} accuracy was checked every {2 hours} to keep within \timeform{2"}. The intensity calibration was applied by observing {W28 [$\alpha_{\rm J2000} =\timeform{18h00m30.426s}, \delta_{\rm J2000} = \timeform{-24D03'58.474"}$] and W44 [$\alpha_{\rm J2000} =\timeform{18h53m18.499s}, \delta_{\rm J2000} = \timeform{-01D14'56.655"}$]}. The final data cube has a beam size of $\sim$ \timeform{28"} and the rms noise level was $\sim$ 0.15 K ch$^{-1}$ for $^{12}$CO $(J =$ 3--2) and $\sim$ 0.10 K ch$^{-1}$ for $^{13}$CO $(J =$ 3--2) with a velocity resolution of 0.3 km s$^{-1}$. {We used the off position as $(l,b)=$(\timeform{352.2D}, \timeform{2.133D}) for the observations with NANTEN2 and ASTE.}
  
  We also used $^{12}$CO ($J=$ 1--0) data cube obtained with the NANTEN 4 m telescope, which was published in Mizuno \& Fukui (2004) and Takeuchi et al. (2010). The data cube has a beam size of \timeform{2.6'} (FWHM) {at a \timeform{4'} grid spacing} and the rms noise revel is $\sim$ 0.45 K ch$^{-1}$ with a velocity resolution of 0.65 km s$^{-1}$.

\section{Results}
\subsection{$^{12}$CO $J=$ 1--0 distribution in a large scale}

The two H\,\emissiontype{II} regions {NGC 6334} and {NGC 6357} are outstanding by high-mass star formation and emit strong infrared and molecular emissions {in} Figures 1 and 2. Enhanced $^{12}$CO emission is also seen toward two places at $(l,b)=$(\timeform{350.6D}, \timeform{1.0D}) and (\timeform{351.8D}--\timeform{352.2D}, \timeform{0.7D}) within a range of $b= $\timeform{0.0D}--\timeform{1.5D}; the former one corresponds to another high-mass star forming region GM 24 ({Gyul$'$Budagyan \& Magakian 1977}), while the other has no known star forming region. We tentatively denominate the latter {MC351.9+0.7}.

{Figures 3 and 4} show longitude-latitude diagrams and longitude-velocity diagrams of $^{12}$CO ($J=$ 1--0, and $J=$ 2--1) emissions, which cover {a region} smaller than Figure 1 toward the two H\,\emissiontype{II} regions. We find the main CO component is peaked at $-4$ km s$^{-1}$ and the secondary component at $-16$ km s$^{-1}$ (the blue-shifted component{, hereafter}). Their peak velocities are nearly constant over the region from $l=$ \timeform{350.5D} to \timeform{353.5D} degrees. The main component was described in the literature (PT08; e.g., Kraemer and Jackson 2009). The blue-shifted component which is weak in intensity ($T_{\rm MB}\sim$ 0.8 K ) was discovered and mapped for the first time in the present work. The narrow component peaked at $+6$ km s$^{-1}$ is the local gas near the Sun, and we find another component at $-25$ km s$^{-1}$ toward $l=$ \timeform{350.9D} to \timeform{351.4D} degrees (in $J=$ 1--0) and $l=$ \timeform{350.5D} to \timeform{352.3D} degrees (in $J=$ 2--1), 
which is more distant than {NGC 6334} {if a kinematic distance is estimated from the Galactic rotation model.} 
In {Figures 3b and 4b} we find bridge features which connect the main component and the blue-shifted component in the following three regions, $l=$ \timeform{351.0D}-- \timeform{351.5D} ({NGC 6334}), $l=$ \timeform{351.8D}--\timeform{352.0D} ({MC351.9+0.7}), and $l=$ \timeform{352.7D}--\timeform{353.3D} ({NGC 6357}). 

{Figure 5 shows the $^{12}$CO intensity line ratio of the $J=$ 2--1 emission to the $J=$ 1--0 emission in the $l$--$b$ and velocity--$l$ diagrams. {We convolved the beam size of the $^{12}$CO ($J=$2--1) to \timeform{180"}, which is the final beam size of the $^{12}$CO ($J=$1--0) data. {The clipping levels of Figure 5(a) and Figure 5(b) are $9\sigma$ (25 K km s$^{-1}$) and 3$\sigma$ (0.3 K degree), respectively.} The ratio is enhanced to be more than 0.8 in the regions of GM 24, {NGC 6334}, and {NGC 6357} in the main component. We also note the bridge features toward NGC6344 and {NGC 6357} shows enhancement of the ratio at 8--10 km s$^{-1}$. The high ratios suggest that the gas is in a high-excitation state {due to heating} by the high-mass stars. }

\subsection{{NGC 6334}}
{Figure 6} shows a $^{12}$CO $J=$ 2--1 distribution of the main component of the NGC 6334 molecular cloud integrated between {$-12$ km s$^{-1}$ and $2$ km s$^{-1}$.} {Figure 7} shows the $^{12}$CO distribution for the blue-shifted component integrated between {$-20$ km s$^{-1}$ and $-12$ km s$^{-1}$}. {Figure 8} shows an overlay of the two components{, which indicates a good correspondence between the two velocity components}.

Figure 9a shows the $^{13}$CO $J=$ 3-2 distribution of NGC 6334. {We identified five $^{13}$CO clumps {toward the far-infrared sub-regions except for II} as listed in Table 2. The clump boundary is defined {at a} 75\% level of the peak integrated intensity. We derived the physical parameters {using the $^{13}$CO $J =$3--2 emission (Buckle et al. 2010) under an assumption of the Local Thermal Equilibrium (LTE).}} Figure 9b shows the radio continuum distribution, {which traces the free-free emission from the H II region}{,} and indicates that sub-regions II, III and IV are coincident with the radio peaks having compact sizes of 1-a few arcmin. For I, I(N) and V nearest radio continuum sources show positional {offsets} from the far-infrared peaks.  We also note that the extended radio emission is seen toward the north of the {NGC 6334} filament {at {Galactic} latitude higher than 0.8 degree}, which {is} a sign of an H\,\emissiontype{II} expanding flow from the regions of {NGC 6334} II and III {to outward} of the Galactic plane.


{Figure 10 shows details of the bridges in NGC 6334 in the $^{12}$CO $J=$2--1 emission. Figure 10a shows spatial distribution of the bridges and indicates peaks toward I, I(N), and V along with weaker extended features. Figures 10b and 10c show line profiles toward {I(N) and I}, and V, respectively, averaged in the two boxes in Figure 10a. The outflow wings in I(N) and I are compact, less than 0.5 pc in size, as shown by two black circles in Figure 10a and are much smaller than the extent of the bridges of $\sim$ 10 pc. We will present discussion on the bridges in Section 4. }

{In Figure 7, we find an extended CO component at $(l,b) = (\timeform{351.0D}$ -- $\timeform{351.8D}, \timeform{0.7D}$ -- $\timeform{1.0D})$, which does not show strong signs of interaction with the main cloud (Figures 8 and 10). It remains as a future task to clarify if the component is part of the interacting gas.}

\subsection{{NGC 6357}}

In {Figures 11 and 12} we find {the patchy molecular cloud toward} {NGC 6357} is extended in the north-south from $b=$ \timeform{0.2D} to \timeform{1.4D} with a central molecular cavity having a $\sim$ 3--5pc radius. The O stars including Pismis 24 lie in the cavity wall. The cavity {has a linewidth of less than 5 km s$^{-1}$ with} no sign of {strong} expansion (see {Figures 3b and 4b}) and may not be simply understood as created by {an effect} of the O stars. In the south below $b=$\timeform{0.8D} we find a clump, part of the main component toward $(l, b)=$(\timeform{353D},\timeform{0.5D}), having a size of (\timeform{0.2D} $\times$ \timeform{0.8D}) elongated with a position angle of 140 degrees. This clump, we tentatively denominate {MC 353.0-0.5} hereafter, is not forming O stars, and remain unionized. 
{{Figure 13}, an overlay of the two velocity components, presents that the {peak of MC 353.0-0.5} shows complementary distribution with an intensity depression in the blue-shifted component. We also find other peaks of the main component correspond to intensity depressions of the blue-shifted component. This complementary distribution as well as the bridge {suggests} physical association of the two velocity components.

{Figure 14 shows details of the bridges in NGC 6357 in the $^{12}$CO $J=$2--1 emission. Figure 14a shows spatial distribution and {indicates} two peaks toward the south of the O stars and along the bubble in the southwest. Figures 14b and 14c show typical line profiles of $^{12}$CO $J=$2--1 toward the peak close to the O stars and another profile in the east of the O stars, which are averaged in the two boxes in Figure 14a, respectively. We develop further discussion on the bridges in Section 4.}

\section{Discussion}

The present study showed that the molecular gas toward the H\,\emissiontype{II} regions {NGC 6334} and {NGC 6357} has two velocity components at {$-4$ km s$^{-1}$} (the main component) and {$-15$ km s$^{-1}$} (the blue-shifted component). The continuity of the main cloud over at least 3 degrees along the plane suggests that the two H\,\emissiontype{II} regions and GM 24 are physically related and have their common parental molecular gas which is elongated {along the plane.} Recent determination {indicates distances} of {NGC 6334} and {NGC 6357} to be around 1.7 kpc (Russeil et al. 2010; see also PT08). At the distance, 3 degrees corresponds to $\sim100$ pc.

The two velocity components seem to be connected in velocity by the bridge features toward {{NGC 6334} and {NGC 6357}}. In addition, the two components show spatially-correlated intensity distributions in spite of the velocity separation; in {NGC 6334} the intensity distributions of the two velocity components are similar with each other, and in {NGC 6357}, {MC353.0-0.5 and its surrounding gas in the red-shifted component {show}} complementary distribution 
{with the intensity depression in the blue-shifted component}. These two observational signatures, the bridge and the spatially correlated distributions, suggest that the two molecular components are physically associated with each other. It has been 
{suggested} that the two signatures are characteristic to high-mass star formation triggered by cloud-cloud collision ({Fukui et al. {2018a}), whereas the positive {intensity correlation between the two velocity components} in {NGC 6334} is a new
 {aspect} in the sample of cloud-cloud collision obtained so far. {We will {discuss these correlations} in the following.
}} 

\subsection{{NGC 6334}}
Molecular clouds in the mini-starburst H\,\emissiontype{II} region {NGC 6334} {show} evidence for cloud-cloud collision which triggered extended O star formation on a 10 pc scale.
The main component toward {NGC 6334} is a filamentary cloud. {The blue-shifted component is weaker than the main component by a factor of more than 10 {at the peak position in Figure 8}.} 
Figure 9a shows the $^{13}$CO $J=$ 3--2 distribution of {NGC 6334} integrated for a velocity range from $-22.5$ to 4 km s$^{-1}$. {NGC 6334} I shows the strongest $^{13}$CO emission while {NGC 6334}II has the weakest $^{13}$CO emission. Figure 9b shows distribution of radio continuum emission obtained with {the Molonglo Observatory Synthesis Telescope (MOST) (Green 1997, Green et al. 1999)}. II shows the strongest radio continuum emission and the weakest $^{13}$CO emission. I(N) and I are associated with the most intense $^{13}$CO emission and their radio continuum emission is weakest (Figure 9). This suggests a trend that older sources having H\,\emissiontype{II} regions show less $^{13}$CO emission probably due to stronger ionization.
The molecular column density of the $^{13}$CO $J=$ 3--2 peaks are calculated from the $^{13}$CO $J=$ 3--2 intensity by assuming LTE (Local thermodynamical equilibrium) of $T_{\rm ex}= 30-40$ K, which is estimated from the $^{12}$ CO peak intensity.

The two molecular components toward {NGC 6334} show similar two peaks at $l=$ \timeform{351.2D} and \timeform{351.4D} at $b \sim$ \timeform{0.65D} {in Figure 8}. This is a property not found in the previous observations in colliding clouds which usually show complementary distribution of two velocity components. The complementary distribution results from a cavity produced in a large cloud by a small cloud according to numerical studies (Habe and Ohta 1992 ; Anathpindika 2010). The cavity is produced until formation of O star(s) which destructs the ambient gas. If the gas in one of the colliding clouds is very dense and form O star(s) soon after the onset of the collision, a cavity is not necessarily produced when O star(s) form (cf. Habe and Ohta 1992). The colliding gas of two velocities then shares intensity enhancements in common directions as a result of shock or radiative heating, providing a positive intensity correlation. The very active O star formation in {NGC 6334} may be consistent with such high density of the main cloud.

We also note that the feedback by formed O stars may be blowing out part of the molecular gas as found at $l =$  \timeform{351.3D}, over a velocity range of 20 km s$^{-1}$ which covers the main cloud and the blue-shifted cloud. This suggests powerful feedback of the O stars. In the peripheral part of the {NGC 6334} cloud we see some extended H\,\emissiontype{II} regions ionized by 13 O/early B stars including N17, N18, N19, N26, N27, N28, N29, N30, N31, N32 and N33 (or in nomenclature GUM 61, GUM 62, GUM 63, GUM 64, and H\,\emissiontype{II} 351.2+0.5 etc. in Table 1 of PT08). These H\,\emissiontype{II} regions are extended with a size of several arcmin and show weak CO emission, suggesting that they are in a more advanced evolutionary stage than the compact heavily obscured regions in the dense main cloud (e.g., Figure 9). They may be former-generation stars formed by more extended cloud-cloud collision, whereas we are not allowed to trace such collision because of dispersal of the parent molecular gas toward the extended H\,\emissiontype{II} regions.

{According to the present scenario, the blue-shifted component lies on the far side and is in collision with the far-side of the main cloud. This configuration predicts high obscuration in the young stars due to the main component, which is consistent with the large optical obscuration toward the O stars in {NGC 6334} where ionization is not yet significant. H\,\emissiontype{I} absorption offers possibly a constraint on the location of the components on the line of sight and is a subject in a future detailed analysis.}

\subsection{{NGC 6357}}
It seems that the two clouds in {NGC 6357} are more heavily dispersed by the ionization than in {NGC 6334}, possibly due to {its older age} than {NGC 6334}. The highest mass stars in the region are the cluster Pismis 24 (Massi et al. 2015). This may be consistent with the large cavity of $\sim$ 3 -- 5 pc radius in the patchy main cloud created by their strong feedback. The age of Pismis 24 is estimated to be {1 Myrs to 7 Myrs (Massi et al. 2015; Lima et al. 2014; Fang et al. 2012; Gvaramadze et al. 2011)}. 

{{Figure 13} shows that complementary distribution between the two velocity components is seen {at} $b \leq$\timeform{0.6D}. The bridge feature and the complementary distribution at the base of {NGC 6357} support the collision. {The NGC 6357} cloud seems to be compressed as shell-like distribution and we did not see clear complementary signatures at galactic latitude above 0.6 degrees. The two velocity components may show in part positive correlations with each other at this high galactic latitude ({Figure 13}). This correlation is also consistent with the physical association between the two velocity components, whereas the origin of the positive correlation may be different from {NGC 6334}. The distribution with a large shell is probably a result of strong stellar feedback by the extremely massive stars in Pismis 24 etc., possibly including some supernova explosions of the most massive member stars. The correlation between the two components may be ascribed to that the two components originated in common gas and were accelerated to different velocity. We are limited in tracing back the star formation history into further details in the more evolved {NGC 6357} than in {NGC 6334}. }

{Figure 14 shows an overlay of the distribution of the bridge features in NGC 6357. The strong bridge is found in the south of the O stars clustered in Pismis 24 at $(l,b)\sim (\timeform{353.18D}, \timeform{0.84D})$, and the weaker extended bridge is found in the south at $(l,b)=($\timeform{353.0D}--\timeform{353.2D}, \timeform{0.6D}--\timeform{0.8D}$)$.The former is likely a remnant of the collision toward the O stars and the latter is not related to O star formation.}

{As an alternative interpretation of the bridges, gas acceleration by the O stars may have created the bridge features. In order to test the possibility we examined two figures, Figures 10 and 14, which present distribution and the typical line profiles of the bridges in NGC 6334 and NGC 6357.
The distribution of the bridge features show two peaks toward the three younger sources {I(N) and I}, and V (Figure 10a). The outflow shows wing intensity of {$7$--$13$ K} over $5$ km s$^{-1}$ velocity span in two velocity ranges, $-10$ -- $-15$ km s$^{-1}$, $10$--$15$ km s$^{-1}$ ({Qiu et al. 2011}).
{These wing intensities are diluted to be less than 0.3 K by averaging in the boxes in Figure 10a. This is because the outflow wing is small in size and has a small beam filling factor.} {The observed intensity of the bridge in Figures 10b and 10c is much higher than the outflow wing which is beam diluted.} {In addition, the red wings are not found at a 3 sigma noise level of 0.4 K in Figures 10b and 10c.} We conclude that the bridge is not explained by the outflow wings. In NGC 6357 the bridge close to the O stars is found between the two velocity components, the red and blue shifted components, and is not extended beyond the blue-shifted component at velocity smaller than $-20$ km s$^{-1}$ (Figures 4c and 14). If the bridge were the blown-out gas by the O stars, its velocity can extend beyond the blue-shifted component near the O stars where blown-out gas would be most significant. Such velocity distribution is not found in Figures 14b and 14c. The other bridge feature in the southwest in Figure 14a is not associated with O stars and is unlikely due to O stars. In summary, the O stars are not considered to have created the bridge features in the present regions.}

\subsection{{MC351.9+0.7}}
{The bridge feature toward $l=$ \timeform{351.8D}--\timeform{352.0D} is intriguing. The region shows enhanced CO intensity and it is likely that the two velocity components are collisionally interacting without star formation. A possibility is that the initial density of the gas was lower than required to form O stars and no O star formation is taking place in spite of the collision. Alternatively, the collision may be too young, e.g., with timescale less than $10^5$ yrs, to form stars. The Herschel image shows filamentary distribution {with infrared emission from cold dust in Figure 1}. According to numerical simulations of cloud-cloud collision (Inoue and Fukui 2013; Takahira et al. 2014; Balfour et al. 2017), formation of filamentary clouds is predicted in collision and may be applied to the filamentary distribution. This is an interesting issue for a future challenge if collision forms generally filamentary distribution in the shocked interface. It is also interesting to look for any dense clumps in the region which may represent shock compressed dense gas leading to future star formation.}

\subsection{Collision model and parameters}
We summarize the cloud-cloud collision in the present two starbursts that the two molecular components of 100 pc length in the Carina-Sagittarius Arm collided with each other within the recent Myr {based on the present CO results.} Considering the age difference between {NGC 6334} and {NGC 6357}, we assume that the main part of {NGC 6357} collided first with the blue-shifted component and, then, the collision toward {NGC 6334} began and is still continuing. {It is notable that the collision is highly uneven in the sense that one of the clouds is dense (average molecular column density $\sim5\times10^{22}$ cm$^{-2}$) and the other much less dense (average molecular column density $\sim2\times10^{21}$ cm$^{-2}$). This unevenness is one of the reasons why the double-peaked molecular emission, and hence the collision, was not recognized in the previous studies of {NGC 6334} and {NGC 6357}. Previous cases of cloud-cloud collision indicate that such unevenness is quite common in cloud-cloud collision (e.g., RCW 38, Fukui et al. 2016).}
 {The typical collision time scale in NGC 6334 is estimated by a ratio of the clump size (e.g., Figure 6) and the relative velocity, to be $\sim 1$ pc/12 km s$^{-1} \sim 10^5$ yrs. In Table 1, we find that most of the sub regions have an age in a range from $10^4$ to $10^5$ yrs as is consistent with the estimate.The oldest sub region II has an age of $10^6$ yrs and may have been formed prior to the collision, whereas the heavy obscuration of the region makes the age estimate crude at best.}
 
{NGC 6334} shows different ages in a range from $10^5$ yrs and 1 Myrs depending {on location} as summarized in {Table 1}. In the collision scenario, this age difference is possibly interpreted in terms of difference in separation between the colliding cloud surfaces. The infrared sources {I(N) and I} are of the most recent formation possibly due to the largest pre-collision separation. {The O star distribution is extended about 10 pc in NGC 6334. It is a point of concern if the distribution reflects the shape of the colliding clouds. }Super star clusters in the {Galaxy} were studied for cloud-cloud collision, and cloud cloud collision was found in the four cases (Wd2, NGC 3603, RCW 38, DSB[2003]179). These clusters are compact with a size of less than 1 pc. In {NGC 6334} O star formation is more extended over 10 pc than in the super star clusters. The present results {indicate} that the O star extent is caused by the extended cloud-cloud collision over 10 pc, lending support for the conclusion reached by Fukui et al. (2016) that the extent of O star distribution reflects the area of collisional shock compression.

{The older age} of the {NGC 6357} collision makes it difficult to construct a detailed collision picture because of cloud dispersal by the ionization/feedback.  {We roughly estimated the collision age by a ratio of the bubble size and the relative velocity to be 10 pc/12 km s$^{-1} \sim 1$ Myr.} The age is consistent with that the ionization front proceeds at $\sim$ 5 km s$^{-1}$ in the dense environment of high-mass star formation (e.g., Fukui et al. 2017b). The patchy distribution of the clouds in the north of the cloud is also consistent with the age. {The typical age of the bubble is estimated to be 1 Myr from a ratio of the bubble radius and the effective velocity width to be 5 pc/6 km s$^{-1}$ (see Figures 3 and 4).}

{Detailes of the shocked layers is a subject suited to ALMA, and high resolution studies will shed a new and important light on the role of the proposed collision in star formation. In this context {MC351.9+0.7} may be of considerable interest since it may be in a very early stage of cloud-cloud collision involving clump/core formation in the earliest stage of high-mass star formation. }

\section{Conclusions}
{We carried out new CO ($J =$ 1--0, 2--1 and 3--2) observations with NANTEN2 and ASTE in the region of twin Galactic mini-starbursts {NGC 6334} and {NGC 6357}. In {NGC 6334} and {NGC 6357}} we detected two velocity components at $-4$ km s$^{-1}$ and $-16$ km s$^{-1}$ with a $\sim12$ km s$^{-1}$ velocity separation, which are linked by bridge features toward the two starburst regions. One of the two components is weak in CO intensity and was detected by the present high sensitivity {$\sim 0.8$ K in $T_{\rm mb}$}. The two velocity components show spatial correlation with each other and with the O stars; toward {NGC 6334} the two components show intensity peaks and depression commonly toward the young H\,\emissiontype{II} regions, and toward {NGC 6357} we found spatial anti-correlated distribution between the two velocity components along the plane. The intensity ratio distributions between the $^{12}$CO $J = 2$--1 and $J = 1$--0 intensities indicate that the ratio is high toward the main cloud which is actively forming O stars. In addition the bridge in {NGC 6334} shows enhanced ratio, indicating its physical association with the O stars. Based on these results we hypothesize that the two clouds in each region collided with each other during the past few Myr and triggered formation of the two starbursts. In {NGC 6334}, we suggest spatial variation of the colliding epoch due to non-uniform cloud separation in order to explain the age difference among the young O stars, where the latest collision happened within $10^5$ yrs. We conclude that cloud-cloud collision offers a possible explanation of the min-starburst over a 100--pc scale active O star formation, where the number of O stars identified until now is 30 with age spread of $\leq 10^5$ -- $10^6$ yrs (PT08). The findings lend support for the collision as an important mode of O star formation in the Carina-Sagittarius Arm. 

\section*{Acknowledgements}
NANTEN2 is an international collaboration of ten universities: Nagoya University, Osaka Prefecture University, University of Cologne, University of Bonn, Seoul National University, University of Chile, University of New South Wales, Macquarie University, University of Sydney, and Zurich Technical University. 
The ASTE telescope is operated by National Astronomical Observatory of Japan (NAOJ).

{Herschel is an ESA space observatory with science instruments provided by European-led Principal Investigator consortia and with important participation from NASA.}
{The Molonglo Observatory Synthesis Telescope is funded by the Australian Research Council and the University of Sydney.}
The work is financially supported by a Grant-in-Aid for Scientific Research (KAKENHI, No. 15K17607, 15H05694) from MEXT (the Ministry of Education, Culture, Sports, Science and Technology of Japan) and JSPS (Japan Soxiety for the Promotion of Science).


\clearpage

\begin{table}
\tbl{Parameters of the infrared sources in {NGC 6334}$^{\dag}$}{%
\begin{tabular}{ccccc}
\hline
\multicolumn{1}{c}{Name} & Galactic Longitude & Galactic Latitude & Total luminosity &  Age\\
& [degree] & [degree] & [$L_{\odot}$] &  [yr]\\
(1) & (2) & (3) & (4) & (5) \\
\hline
I(N) & \timeform{351.44} & \timeform{0.660} & $4.9\times10^4$ & $<10^4$ \\
I & \timeform{351.42} & \timeform{0.644} & $1.5\times10^5$ & $10^4$ \\
II & \timeform{351.37} & \timeform{0.647} & $2.6\times10^5$ & $10^6$ \\
III & \timeform{351.32} & \timeform{0.663} & $1.8\times10^5$ & $10^5$\\
IV & \timeform{351.24} & \timeform{0.666} & $2.2\times10^5$ & $10^5$ \\
V & \timeform{351.16} & \timeform{0.697} & $1.9\times10^5$ & $10^5$\\
\hline
\end{tabular}}\label{tab:first}
\begin{tabnote}
\footnotemark[] Note. Columns: (1) Name. (2,3) Position. (4) Total luminosity of infrared source. (5) Age of the infrared source.\\
\footnotemark[$\dag$] Reference: Persi \& Tapia (2008) \\
\end{tabnote}
\end{table}

\begin{table}
\tbl{Parameters of the $^{13}$CO $J=3-2$ Clumps in {NGC 6334}}{%
\begin{tabular}{p{2cm}p{2cm}p{2cm}p{2cm}p{2cm}}
\hline
\multicolumn{1}{c}{Name} &  $T_{\rm ex}$ & $N(H_2)$ & Size & $M_{\rm LTE}$ \\
&  [K] & [cm$^{-2}$] & [pc] & [$M_{\odot}$]\\
\multicolumn{1}{c}{(1)} & (2) & (3) & (4) & (5) \\
\hline
\multicolumn{1}{c}{I(N)} & 38 & $4 \times 10^{22}$ & 0.46 & $1 \times 10^3$\\
\multicolumn{1}{c}{I} & 41 & $4 \times 10^{22}$ & 0.53 & $7 \times 10^3$ \\
\multicolumn{1}{c}{II} &  --- &--- & --- & --- \\
\multicolumn{1}{c}{III} & 40 &$2 \times 10^{22}$ & 0.30 & $	1 \times 10^3$\\
\multicolumn{1}{c}{IV} & 41  &$4 \times 10^{22}$ & 0.29 & $4 \times 10^3$\\
\multicolumn{1}{c}{V} &  34 &$3 \times 10^{22}$ & 0.19 & $8 \times 10^2$\\
\hline
\end{tabular}}\label{tab:first}
\begin{tabnote}
\footnotemark[] Note. Columns: (1) Name. (2) Excitation temperature of $^{12}$CO $J=$ 1--0 peak intensity. (3) Average column density of the $^{13}$CO clump {using the abundance ratio X (CO) = [$^{13}$CO]/[H$_2$] $= 1.4\times10^{-6}$ (Frerking et al. 1982) }. (4) Size of the $^{13}$CO clump {assuming spherical shapes is estimated from the equation of $r=\sqrt{S / \pi}$, where $S$ is the area of the clump enclosed by the boundary.} (5) Molecular mass within the clump size with the assumption of LTE. \\
\end{tabnote}
\end{table}

\begin{figure*}[h]
\begin{center}  \includegraphics[width=17cm]{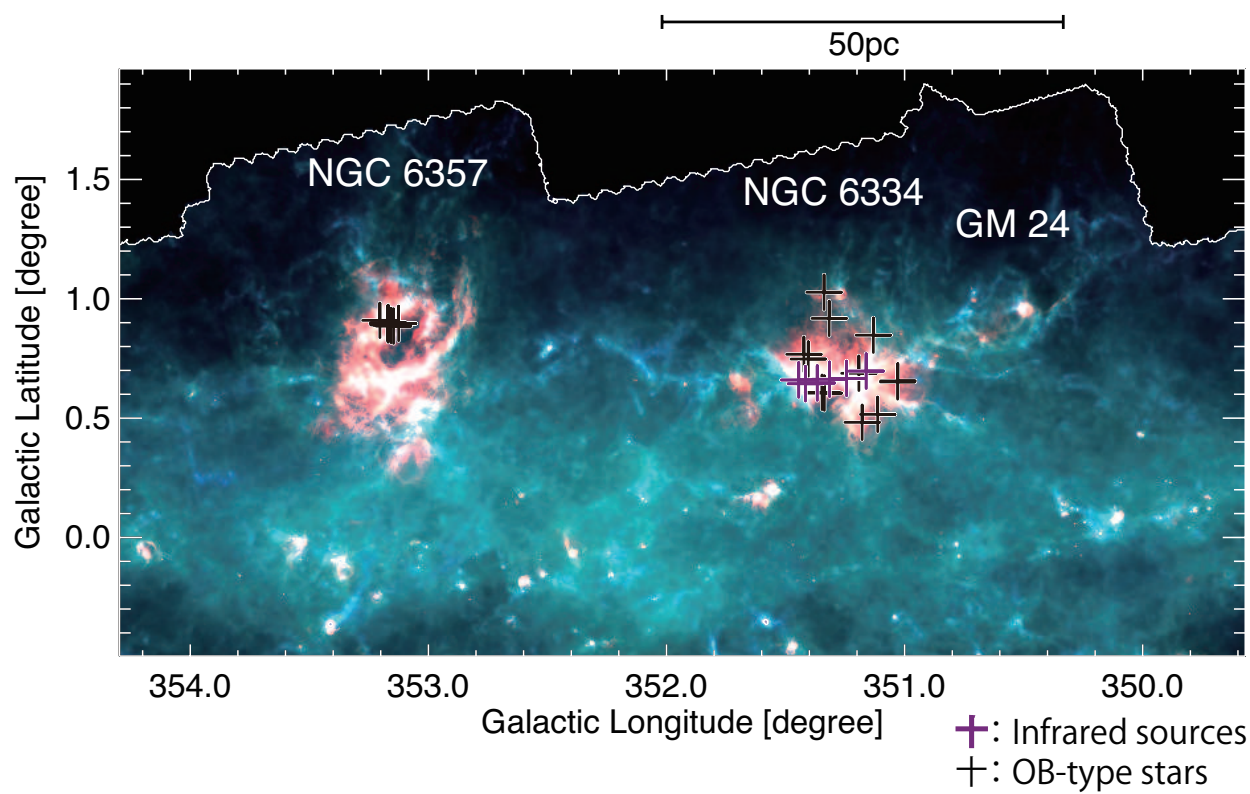}
\end{center}
\caption{Three color image of {NGC 6334}-6357.  Red, green, and blue show the Herschel/PACS 70 $\mu$m, Herschel/SPIRE 250 $\mu$m, Herschel/SPIRE 500 $\mu$m {(Molinari et al. 2010, Pilbratt et al. 2010; Poglitsch et al. 2010; Griffin et al. 2010)}. {{Purple} and black crosses indicate infrared sources (PT08) and OB-type stars (PT08, Fang et al. 2012), respectively.}}\label{.....}
\end{figure*}

\begin{figure*}[h]
\begin{center}  \includegraphics[width=17cm]{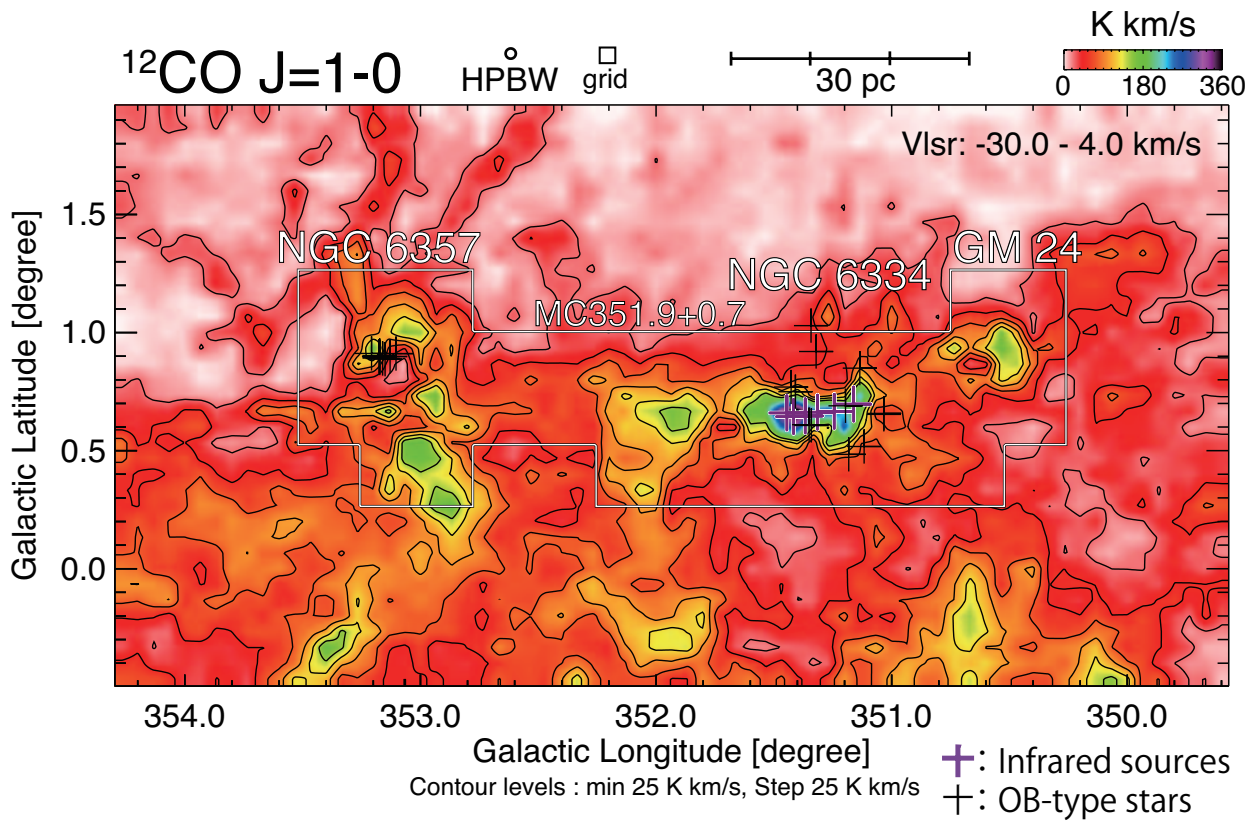}
\end{center}
\caption{Integrated intensity map of $^{12}$CO $J=$ 1--0 obtained with NANTEN. {Purple and black crosses indicate infrared sources (PT08) and OB-type stars (PT08, Fang et al. 2012), respectively. The beam and grid spacing are \timeform{2.6'} and \timeform{4'}, respectively. The integrated velocity range is from $-30$ to $4$ km s$^{-1}$. The $1\sigma$ noise level is $\sim 2.1$ K km s$^{-1}$ for the velocity interval of 34 km s$^{-1}$. The lowest contour and contour intervals are 25 K km s$^{-1}$ and 25 K km s$^{-1}$, respectively. The white line shows the observed area of $^{12}$CO $J= $ 2--1 with NANTEN2.}}\label{.....}
\end{figure*}

\begin{figure*}[h]
\begin{center}  \includegraphics[width=17cm]{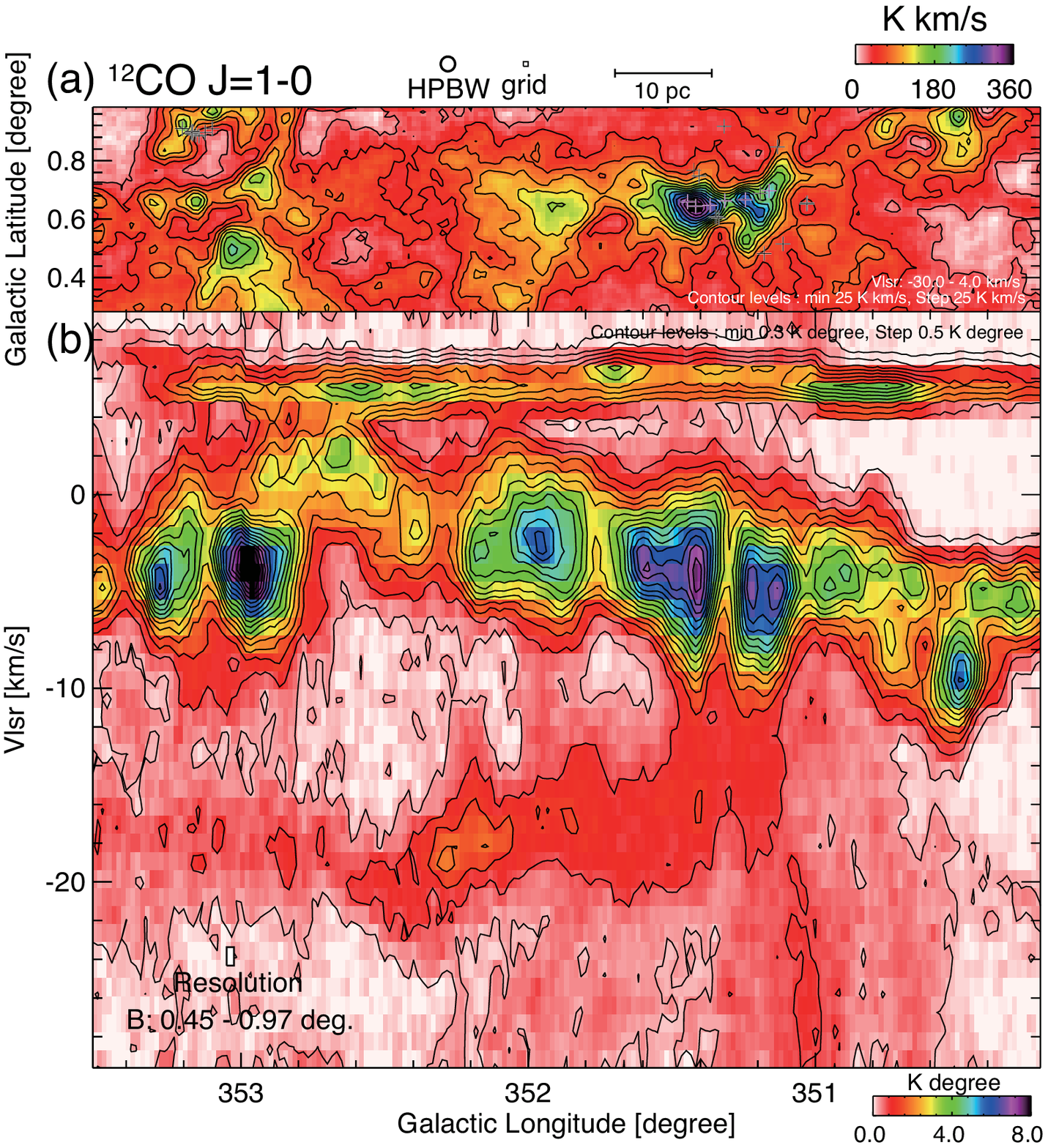}
\end{center}
\caption{(a) Integrated intensity map of $^{12}$CO $J=$1--0 obtained with NANTEN2. {Purple and black crosses indicate infrared sources (PT08) and OB-type stars (PT08, Fang et al. 2012), respectively. The final beam size after convolution and grid spacing are \timeform{180"} and \timeform{60"}, respectively. The integrated velocity range is from $-30$ to $4$ km s$^{-1}$. The $1\sigma$ noise level is $\sim 2.8$ K km s$^{-1}$ for the velocity interval of 34 km s$^{-1}$. The lowest contour and contour intervals are 25 K km s$^{-1}$.} (b) Galactic Longitude-Velocity diagram of $^{12}$CO $J=$1--0. {The integrated latitude range is from $0.45$ to $0.97$ degree. The velocity resolution is smoothed to 1 km s$^{-1}$. The $1\sigma$ noise level is $\sim 0.1$ K degree for the latitude interval of 0.54 degree. The lowest contour and contour intervals are 0.3 K degree and 0.5 K degree, respectively. }}\label{.....}
\end{figure*}

\begin{figure*}[h]
\begin{center}  \includegraphics[width=17cm]{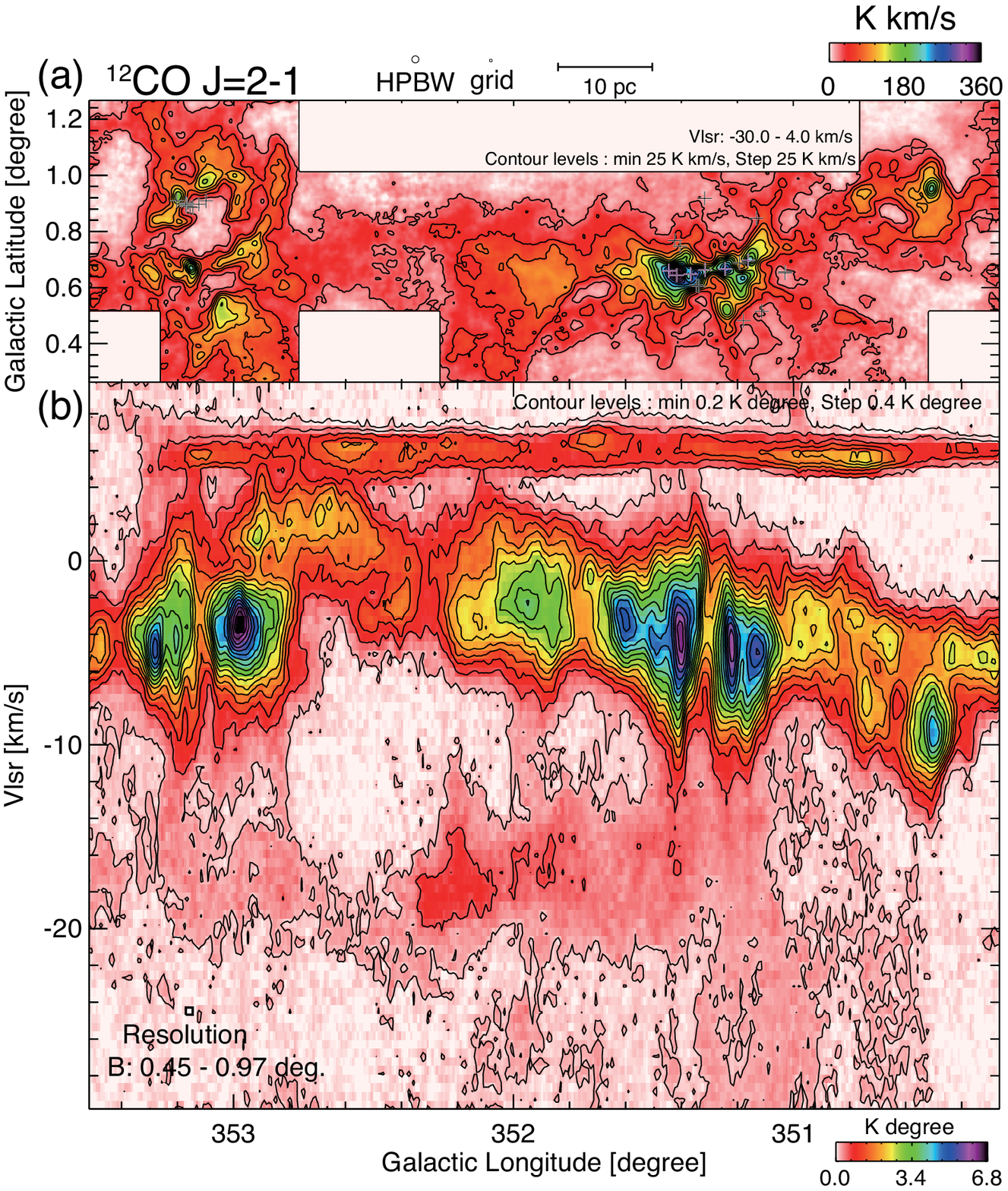}
\end{center}
\caption{(a) Integrated intensity map of $^{12}$CO $J=$2--1 obtained with NANTEN2. {Purple and black crosses indicate infrared sources (PT08) and OB-type stars (PT08, Fang et al. 2012), respectively. The final beam size after convolution and grid spacing are \timeform{90"} and \timeform{30"}, respectively. The integrated velocity range is from $-30$ to $4$ km s$^{-1}$. The $1\sigma$ noise level is $\sim 1.8$ K km s$^{-1}$ for the velocity interval of 34 km s$^{-1}$. The lowest contour and contour intervals are 25 K km s$^{-1}$.}(b) Galactic Longitude-Velocity diagram of 
$^{12}$CO $J=$ 2--1. {The integrated latitude range is from $0.45$ to $0.97$ degree. The velocity resolution is smoothed to 0.5 km s$^{-1}$. The $1\sigma$ noise level is $\sim 0.07$ K degree for the latitude interval of 0.54 degree. The lowest contour and contour intervals are 0.2 K degree and 0.4 K degree, respectively.}}\label{.....}
\end{figure*}

\begin{figure*}[h]
\begin{center}  \includegraphics[width=17cm]{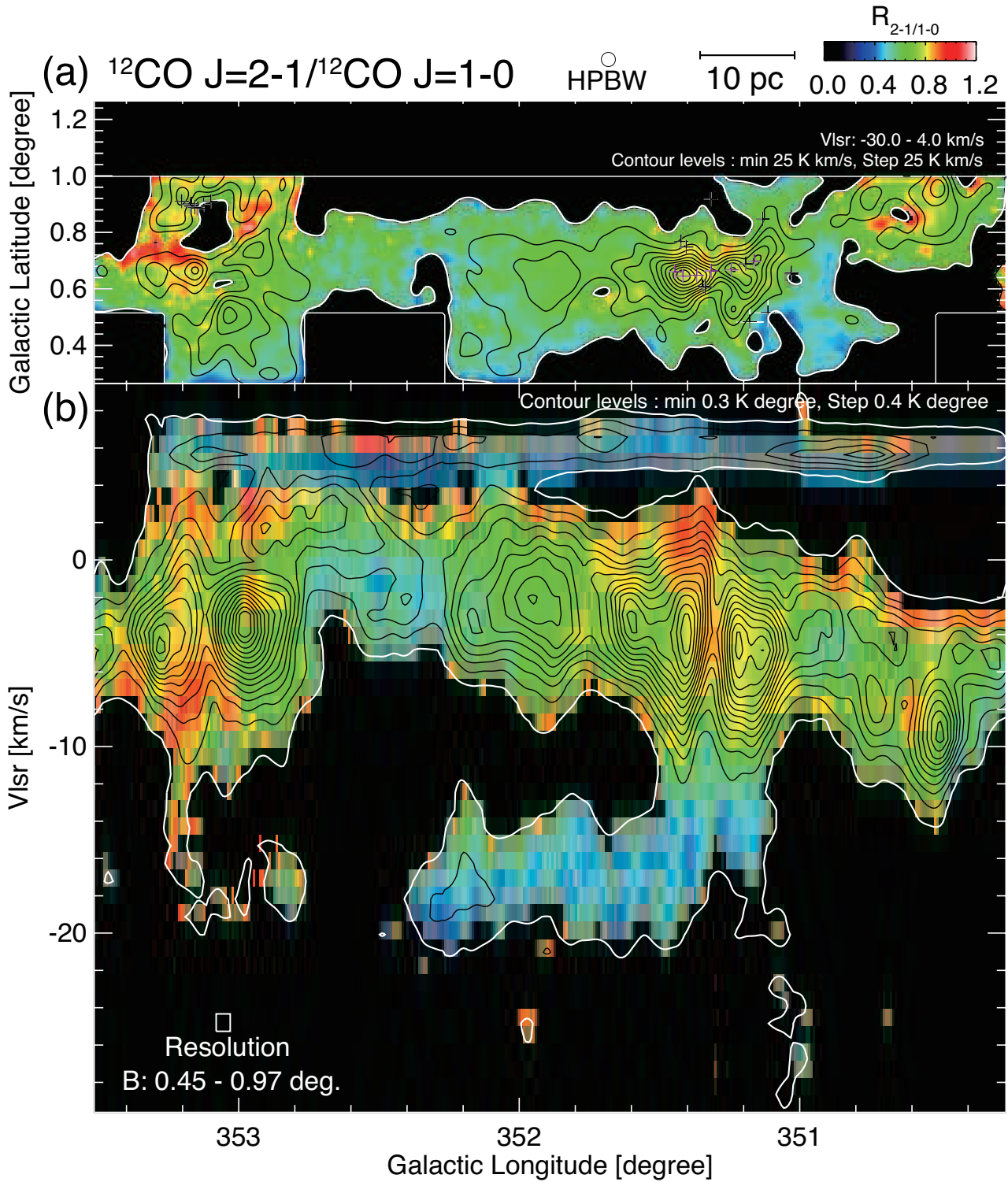}
\end{center}
\caption{(a) Intensity ratio map of $^{12}$CO $J=$ 2--1/$^{12}$CO $J=$ 1--0 with NANTEN2. {Purple and black crosses indicate infrared sources (PT08) and OB-type stars (PT08, Fang et al. 2012), respectively. The final beam size after convolution is \timeform{180"}. The integrated velocity range is from $-30$ to $4$ km s$^{-1}$. The $1\sigma$ noise level is $\sim 2.8$ K km s$^{-1}$ for the velocity interval of 34 km s$^{-1}$. The lowest contour and contour intervals are 25 K km s$^{-1}$ in $^{12}$CO $J=$2--1.} (b) Galactic Longitude-Velocity diagram of 
$^{12}$CO $J=$ 2--1/$^{12}$CO $J=$ 1--0. {The integrated latitude range is from $0.45$ to $0.97$ degree. The velocity resolution is smoothed to 1 km s$^{-1}$. The $1\sigma$ noise level is $\sim 0.1$ K degree for the latitude interval of 0.54 degree. The lowest contour and contour intervals are 0.3 K degree and 0.4 K degree, respectively.}}\label{.....}
\end{figure*}


\begin{figure*}[h]
\begin{center}  \includegraphics[width=17cm]{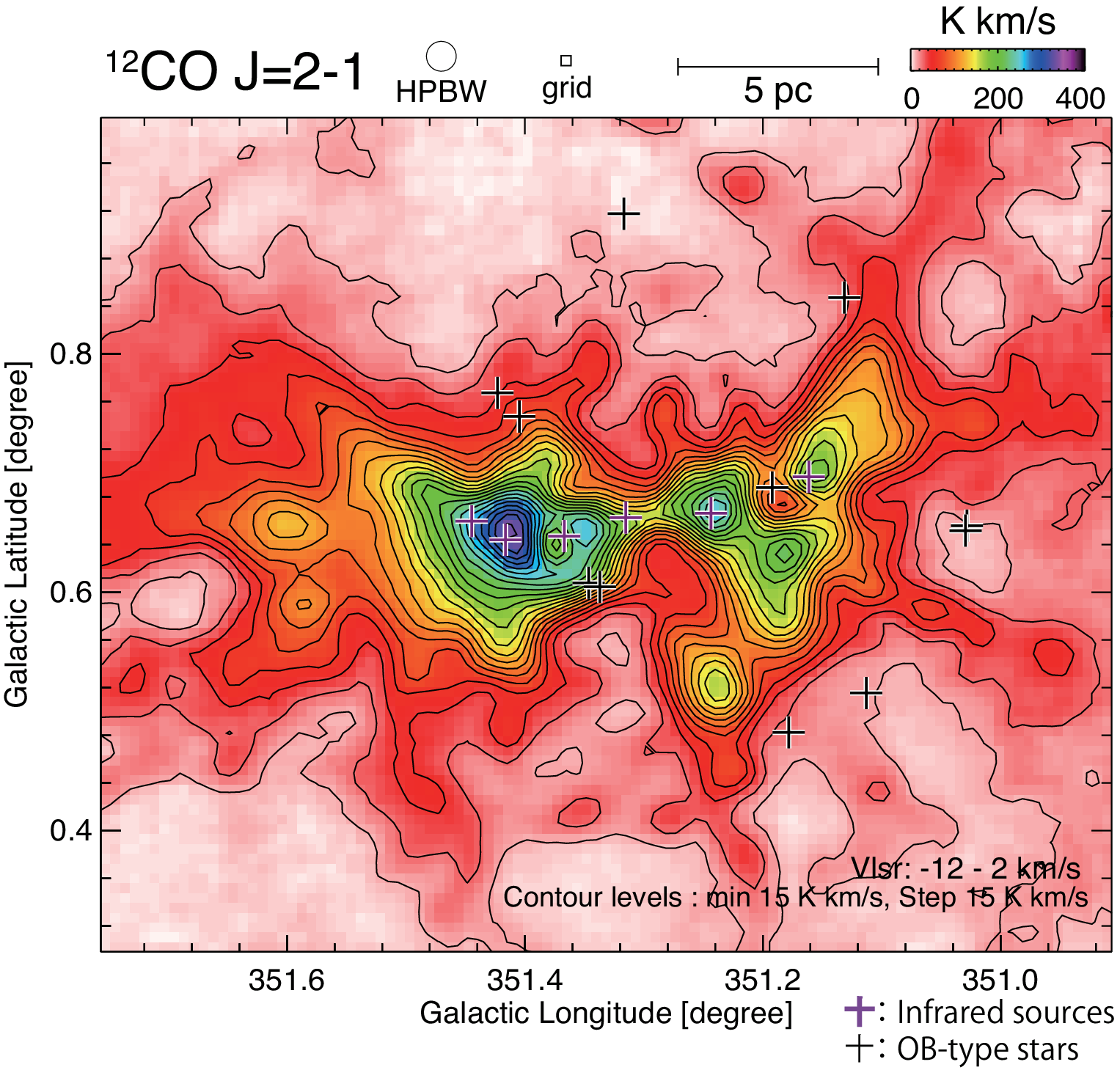}
\end{center}
\caption{Integrated intensity map of $^{12}$CO $J=$ 2--1 obtained with NANTEN2. {Purple and black crosses indicate infrared sources (PT08) and OB-type stars (PT08, Fang et al. 2012), respectively. The final beam size after convolution and grid spacing are \timeform{90"} and \timeform{30"}, respectively. The integrated velocity range is from $-12$ to $2$ km s$^{-1}$. The $1\sigma$ noise level is $\sim 1.2$ K km s$^{-1}$ for the velocity interval of 14 km s$^{-1}$. The lowest contour and contour intervals are 15 K km s$^{-1}$}}\label{.....}
\end{figure*}

\begin{figure*}[h]
\begin{center}  \includegraphics[width=17cm]{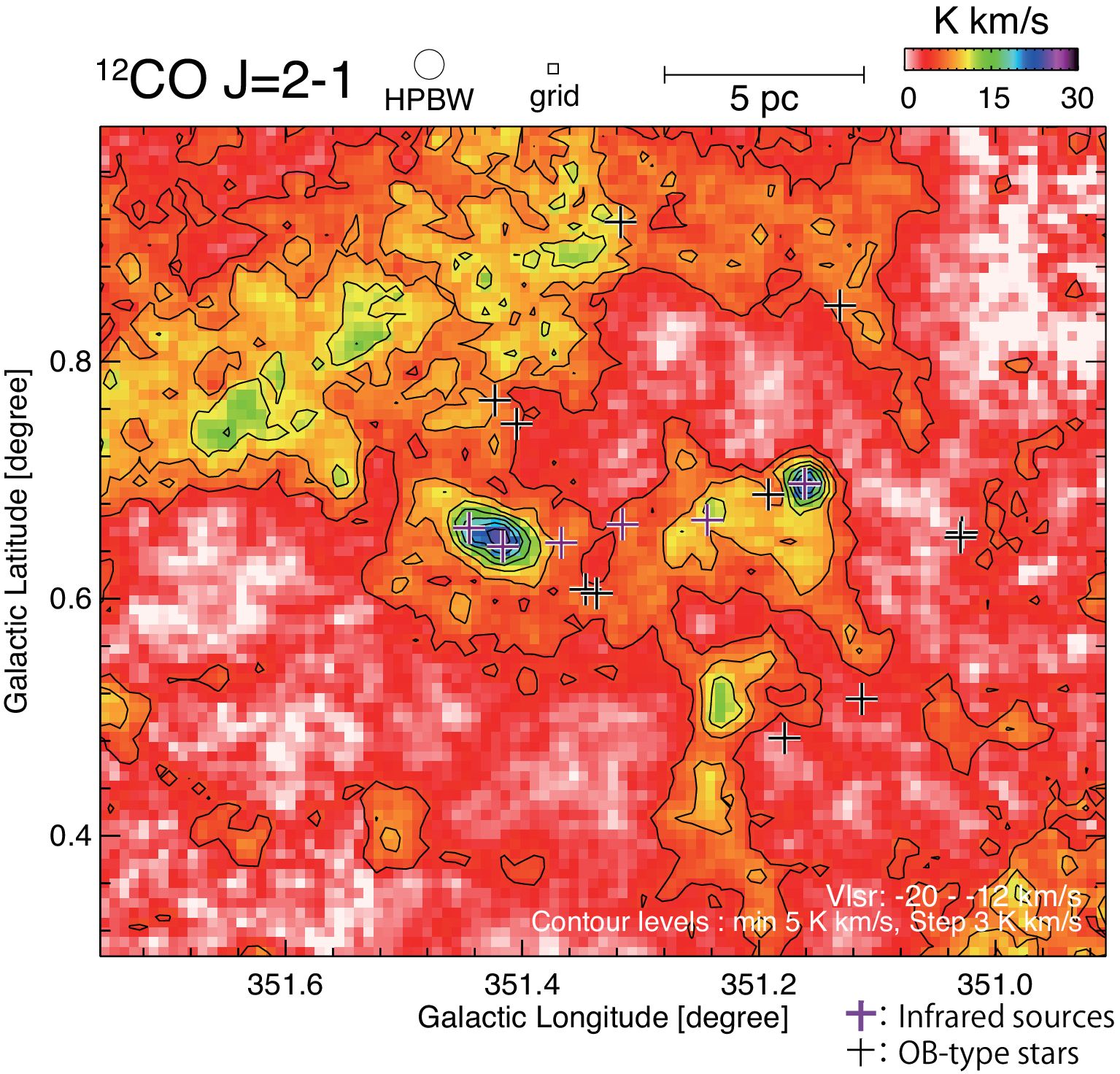}
\end{center}
\caption{Integrated intensity map of $^{12}$CO $J=$ 2--1 obtained with NANTEN2. {Purple and black crosses indicate infrared sources (PT08) and OB-type stars (PT08, Fang et al. 2012), respectively. The final beam size after convolution and grid spacing are \timeform{90"} and \timeform{30"}, respectively. The integrated velocity range is from $-20$ to $-12$ km s$^{-1}$. The $1\sigma$ noise level is $\sim 0.88$ K km s$^{-1}$ for the velocity interval of 8 km s$^{-1}$. The lowest contour and contour intervals are 5 K km s$^{-1}$, and 3 K km s$^{-1}$, respectively.}}\label{.....}
\end{figure*}

\begin{figure*}[h]
\begin{center}  \includegraphics[width=17cm]{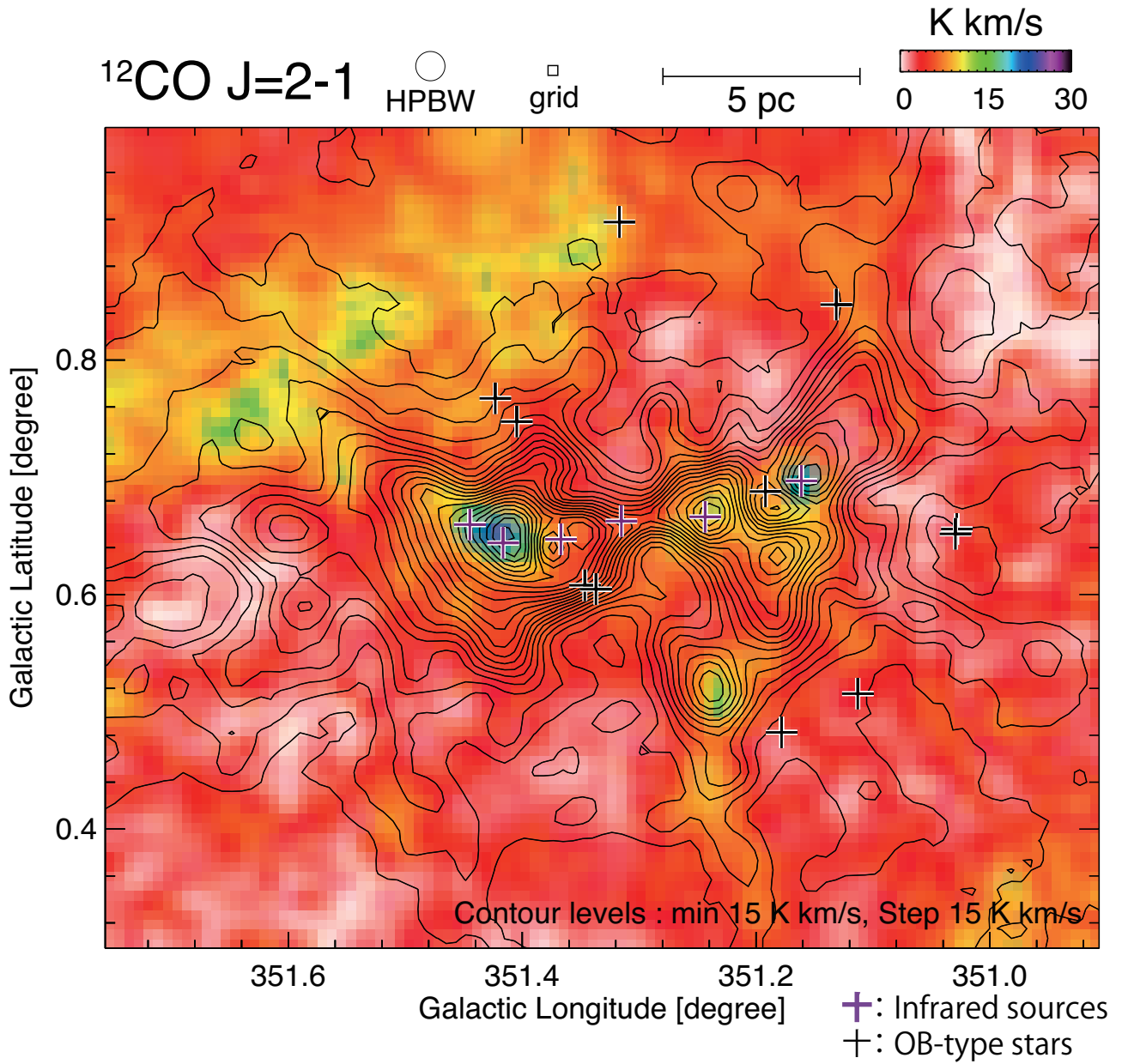}
\end{center}
\caption{Integrated intensity map of $^{12}$CO $J=$ 2--1 obtained with NANTEN2. {Purple and black crosses indicate infrared sources (PT08) and OB-type stars (PT08, Fang et al. 2012), respectively. The final beam size after convolution and grid spacing are \timeform{90"} and \timeform{30"}, respectively. The integrated velocity range of the color map and the contour map is from $-20$ to $-12$ km s$^{-1}$ and $-12$ to $2$ km s$^{-1}$, respectively. The lowest contour and contour intervals are 15 K km s$^{-1}$.}}\label{.....}
\end{figure*}

\begin{figure*}[h]
\begin{center}  \includegraphics[width=17cm]{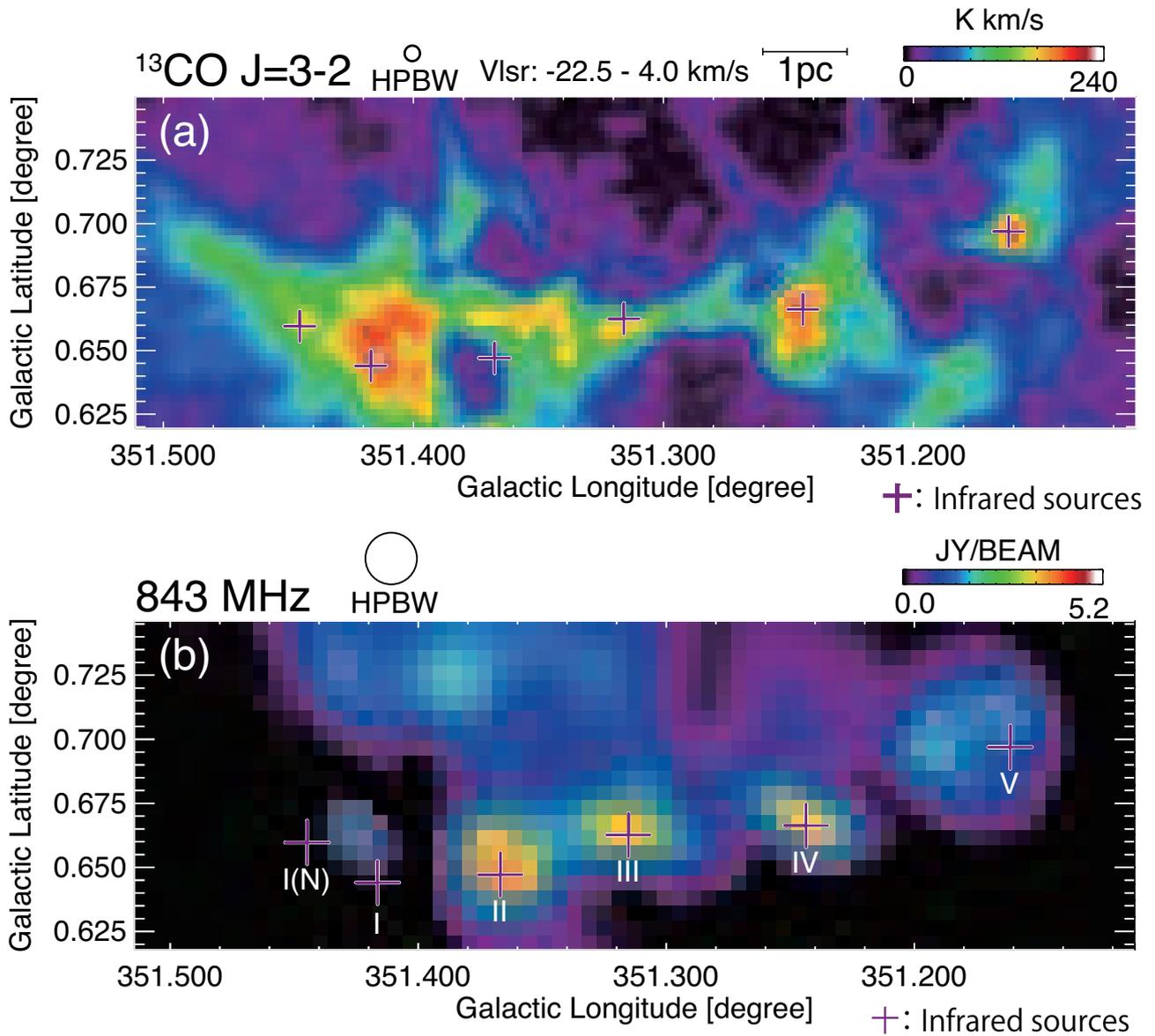}
\end{center}
\caption{(a) Integrated intensity map of {$^{13}$}CO $J=$ 3--2 obtained with ASTE.  {{Purple} crosses indicate infrared sources (PT08). The final beam size after convolution are \timeform{28"}. The integrated velocity range is from $-22.5$ to $4.0$ km s$^{-1}$. The $1\sigma$ noise level is $\sim 0.28$ K km s$^{-1}$ for the velocity interval of 26.5 km s$^{-1}$.} (b) Intensity map of radio continuum ({843 MHz}) obtained with MOST. {{Purple crosses} indicate infrared sources (PT08). The beam size is \timeform{73"}}}\label{.....}
\end{figure*}

\begin{figure*}[h]
\begin{center}  \includegraphics[width=14cm]{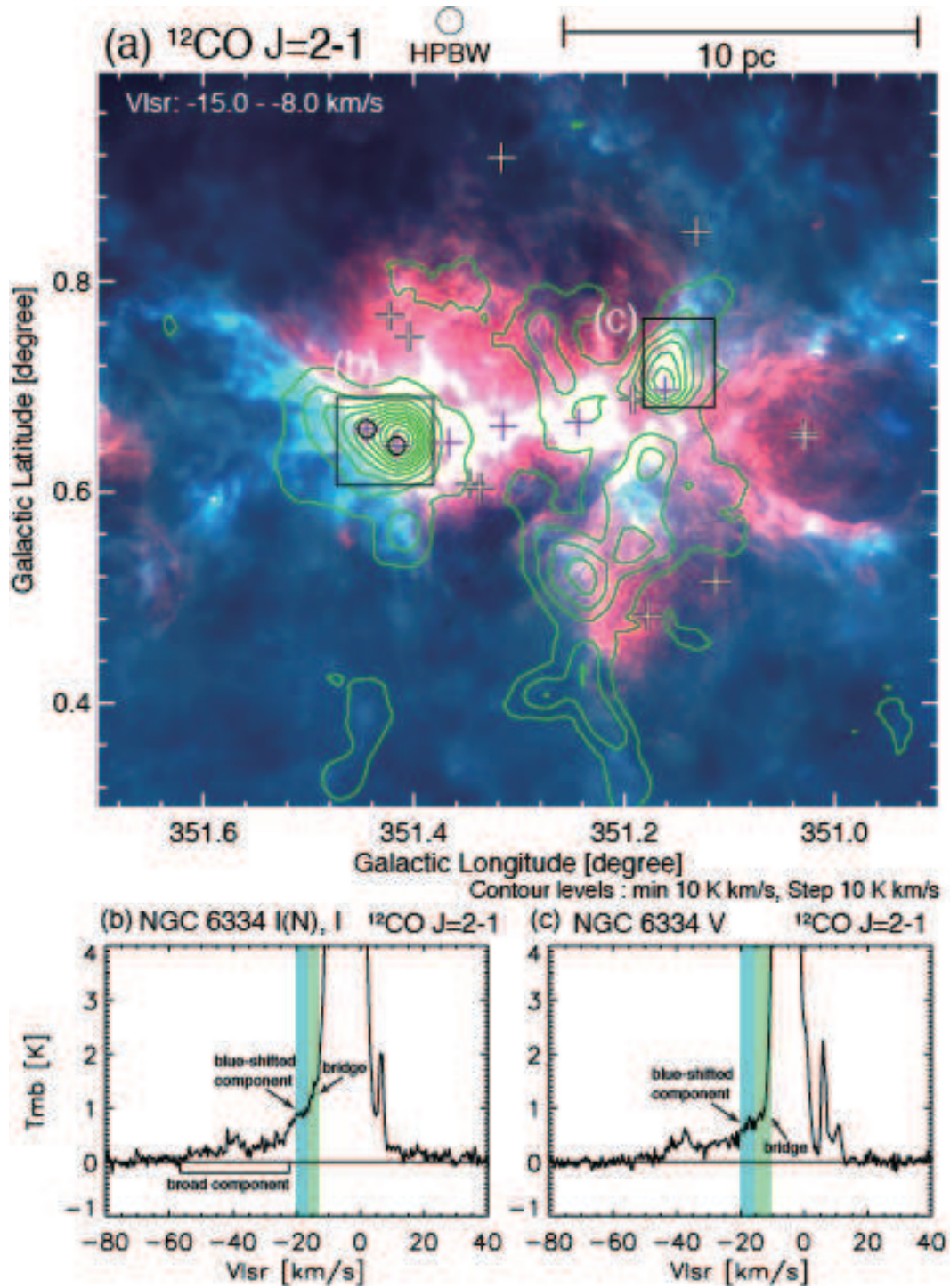}
\end{center}
\caption{{(a) Comparison of the bridging feature (green contour) with Herschel at NGC 6334.  Red, green, and blue show the Herschel/PACS 70 $\mu$m, Herschel/SPIRE 250 $\mu$m, Herschel/SPIRE 500 $\mu$m, respectively. {The green contours show the bridging feature} integrated over a velocity range $-15$ to $-8$ km s$^{-1}$. The two circles toward {I(N) and I} show the maximum extents of the protostellar outflows at five sigmas (McCutcheon et al. 2000). Purple and {white crosses} indicate infrared sources (PT08) and OB-type stars (PT08, Fang et al. 2012), respectively. The {boxes} of (b) and (c) show the averaged area of profiles. (b) The averaged spectra at NGC 6334 I(N) and I. {The box size ($l \times b$) is \timeform{5.5'} $\times$ \timeform{5'}.} (c) The averaged spectra at NGC 6334 V. {The box size ($l \times b$) is \timeform{4'} $\times$ \timeform{5'}.}}}\label{.....}
\end{figure*}

\begin{figure*}[h]
\begin{center}  \includegraphics[width=17cm]{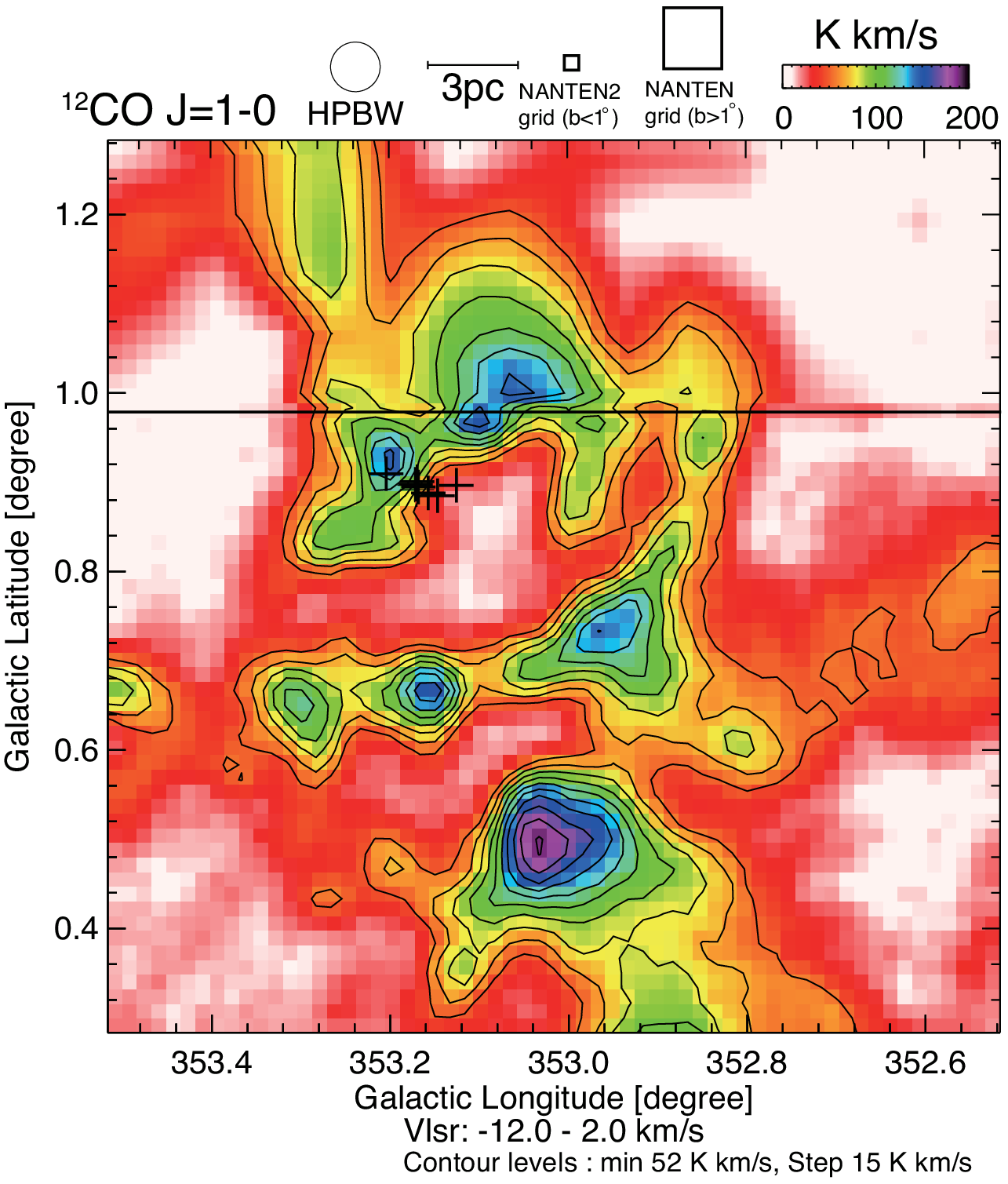}
\end{center}
\caption{Integrated intensity map of $^{12}$CO $J=$ 1--0 obtained with NANTEN and NANTEN2. {The solid line shows the boundary of the two different grid spacing {data sets} from NANTEN and NANTEN2. Black crosses indicate OB-type stars (Fang et al. 2012). The HPBW shows the final beam size of NANTEN2. The integrated velocity range is from $-12$ to $2$ km s$^{-1}$. The $1\sigma$ noise levels of NANTEN and NANTEN2 are $\sim $ 1.4 K km s$^{-1}$  and 1.8 K km s$^{-1}$ for the velocity interval of 14 km s$^{-1}$, respectively. The lowest contour and contour intervals are 52 K km s$^{-1}$ and 15 K km s$^{-1}$, respectively.}}\label{.....}
\end{figure*}

\begin{figure*}[h]
\begin{center}  \includegraphics[width=17cm]{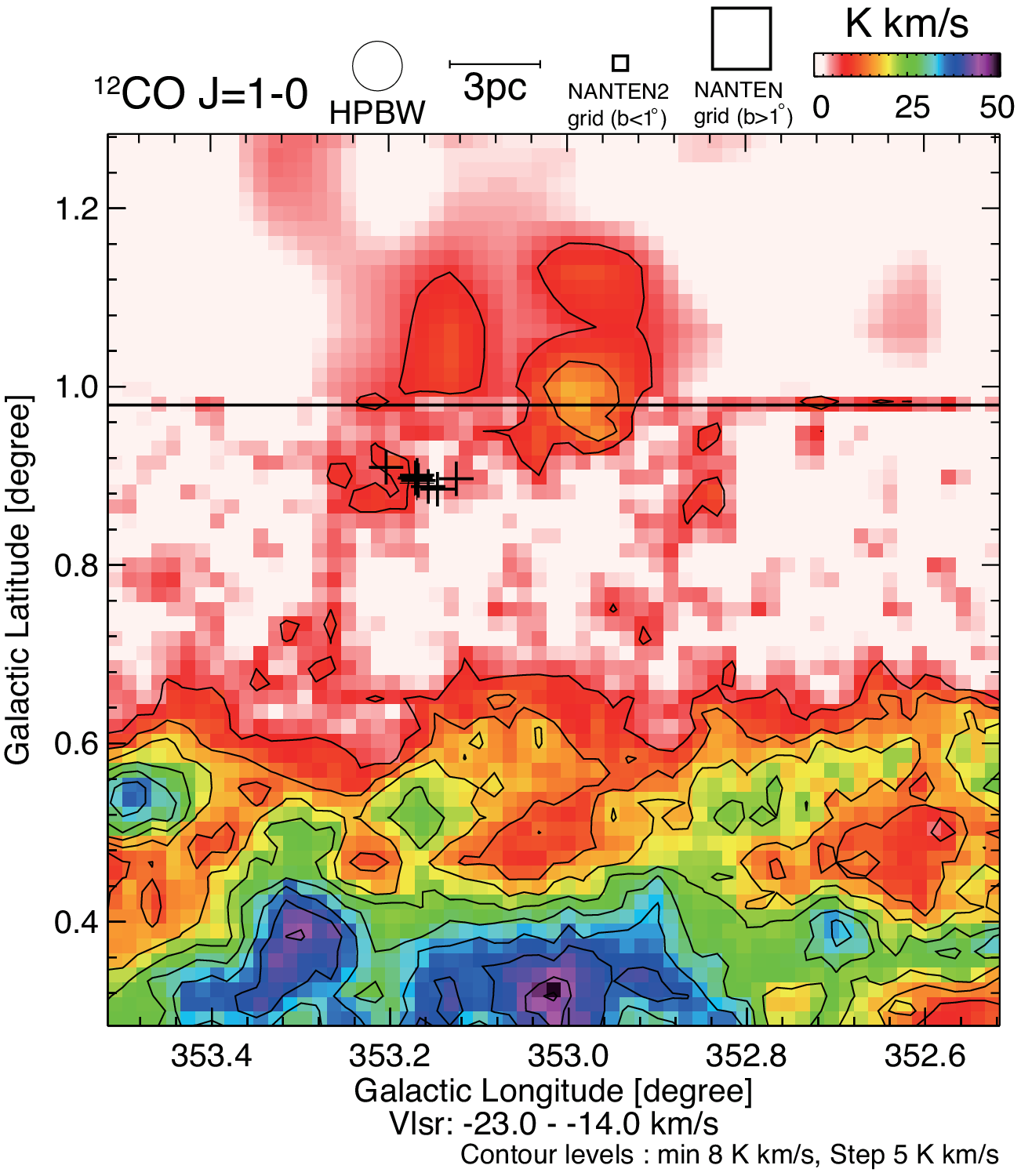}
\end{center}
\caption{Integrated intensity map of $^{12}$CO $J=$ 1--0 obtained with NANTEN and NANTEN2. {The solid line shows the boundary of the two different grid spacing {data sets} from NANTEN and NANTEN2. Black crosses indicate OB-type stars (Fang et al. 2012). The HPBW shows the final beam size of NANTEN2. The integrated velocity range is from $-23$ to $-14$ km s$^{-1}$. The $1\sigma$ noise levels of NANTEN and NANTEN2 are $\sim $ 1.1 K km s$^{-1}$  and 1.4 K km s$^{-1}$ for the velocity interval of 9 km s$^{-1}$, respectively. The lowest contour and contour intervals are 8 K km s$^{-1}$ and 5 K km s$^{-1}$, respectively.}}\label{.....}
\end{figure*}

\begin{figure*}[h]
\begin{center}  \includegraphics[width=17cm]{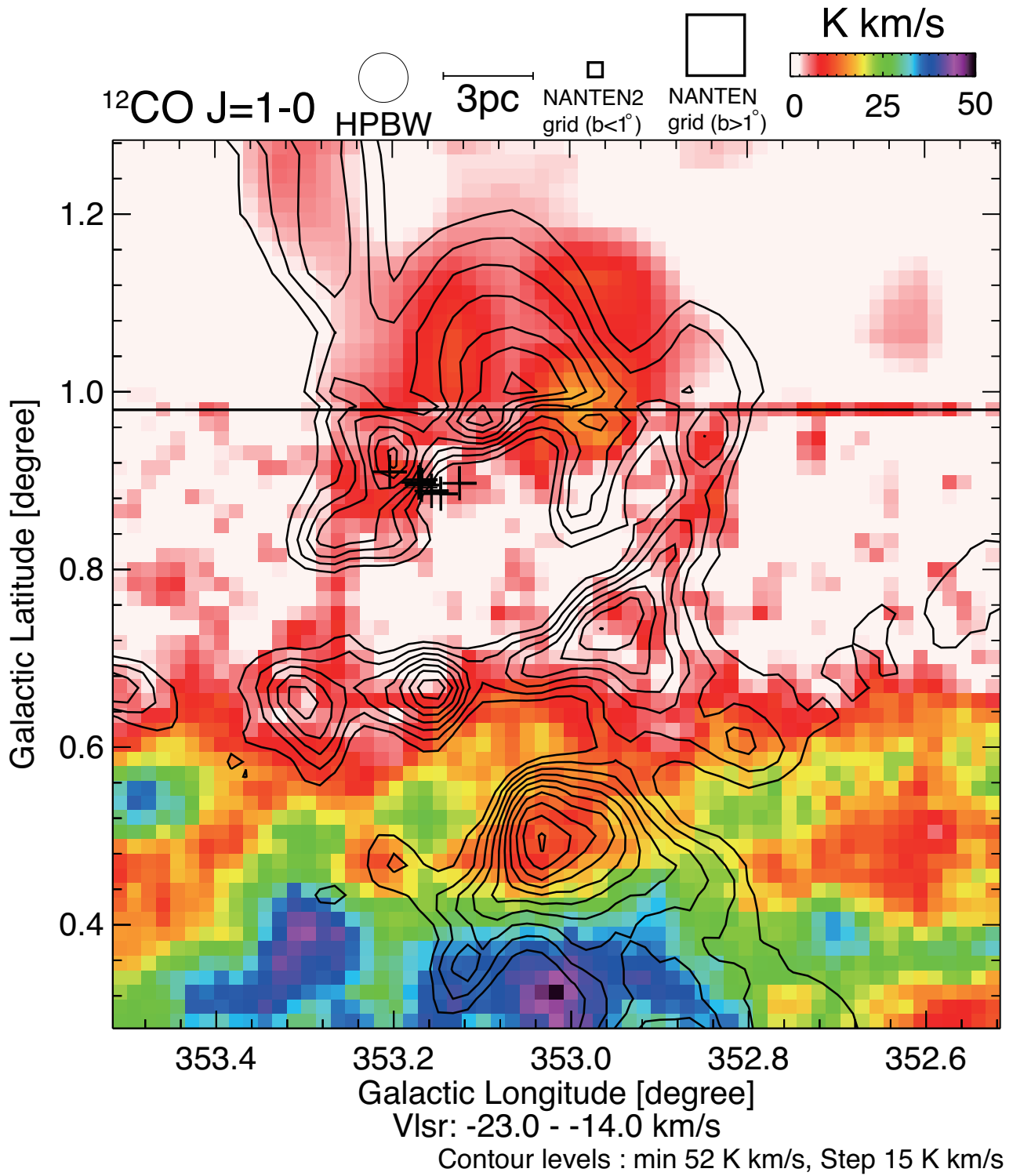}
\end{center}
\caption{Integrated intensity map of $^{12}$CO $J=$ 1--0 obtained with NANTEN and NANTEN2. {The solid line shows the boundary of the two different grid spacing {data sets} from NANTEN and NANTEN2. Black crosses indicate OB-type stars (Fang et al. 2012). The HPBW shows the final beam size of NANTEN2.} The integrated velocity range of the color map is from $-23.0$ to $-14.0$ km s$^{-1}$. The integrated velocity range of the contour map is from -12.0 to 2.0 km s$^{-1}$. }\label{.....}
\end{figure*}

\begin{figure*}[h]
\begin{center}  \includegraphics[width=14cm]{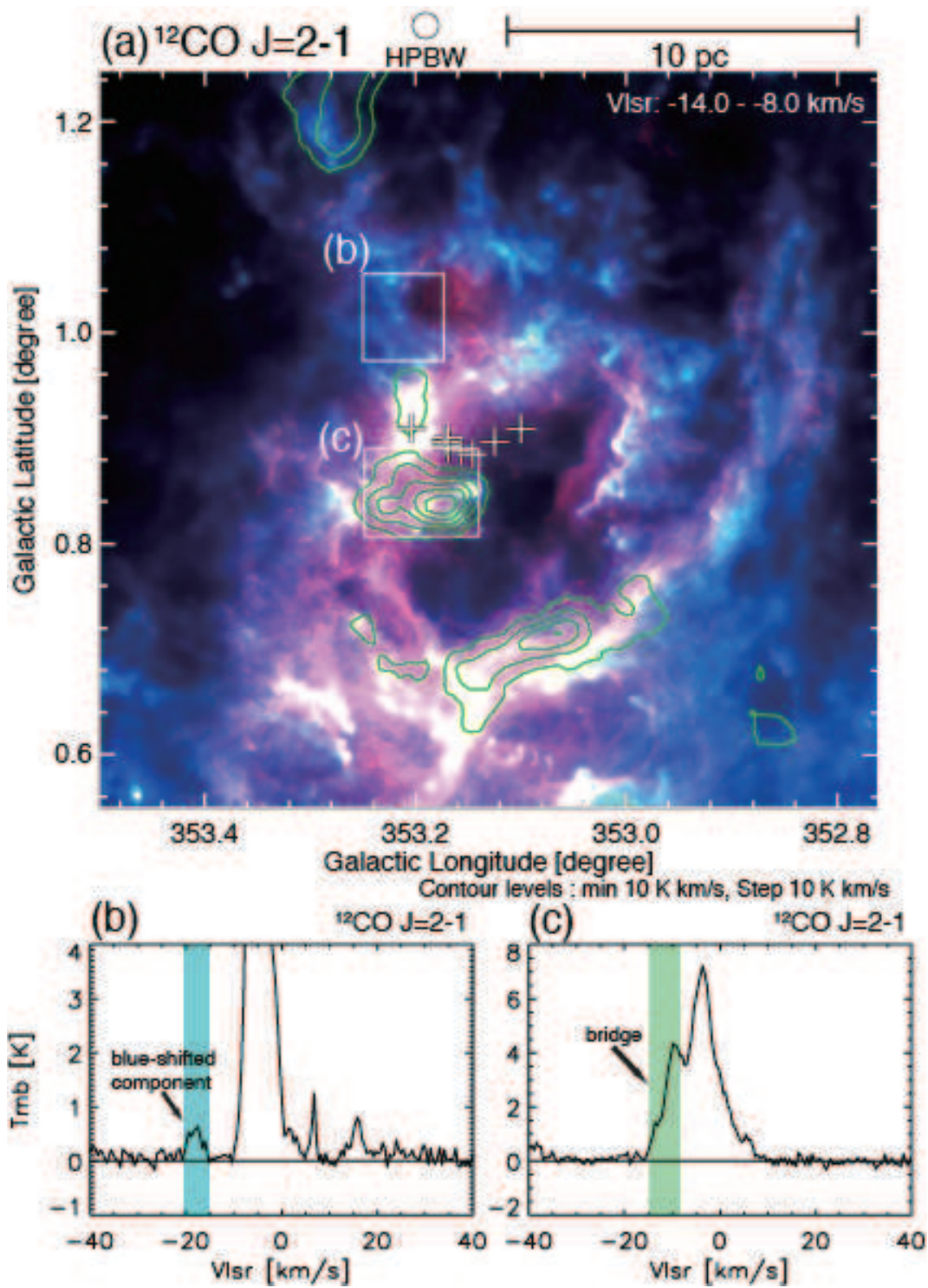}
\end{center}
\caption{{(a) Comparison of the bridging feature (green contour) with Herschel at NGC 6357.  Red, green, and blue show the Herschel/PACS 70 $\mu$m, Herschel/SPIRE 250 $\mu$m, Herschel/SPIRE 500 $\mu$m, respectively. The integrated velocity range is from $-14$ to $-8$ km s$^{-1}$. {White} crosses indicate OB-type stars (Fang et al. 2012). The {boxes} of (b) and (c) show the averaged area of profiles. (b) The averaged spectra {of blue-shifted component. The box size ($l \times b$) is \timeform{5'} $\times$ \timeform{5'}} (c) The averaged spectra {of bridging feature. The box size ($l \times b$) is \timeform{6'} $\times$ \timeform{5'}.}}}\label{.....}
\end{figure*}

\end{document}